# Input Current Sensorless Control for a AC-DC Converter with Unknown Load and Source Voltage Amplitude

Mehdi Tavan and Kamel Sabahi

*Abstract*— Input current estimation is indispensable in the sensorless control algorithms for the problem of power factor compensation (PFC) of an AC-DC boost converter. The system estimator design is challenged by the bilinear form dynamics and uncertain parameters of the system. In this paper, the system dynamics is *immersed* to a proper form by a new filtered transformation. Thanks to the proposed transformation, the input current, input voltage amplitude, and load conductance are globally estimated. The exponential convergent of the estimates is established in normal converter operation. An application of the proposed estimator is presented in conjunction with a well-known dynamic controller.

## I. INTRODUCTION

Fig. 1 shows the single-phase full-bridge boost converter. From this figure, it can be seen that the circuit of the converter benefits from two pair of transistor-diode switches in two legs to form a bidirectional operation. Using a PWM circuit controller, the switches in each leg operate in complementary way. From the Kirchhoff's laws, the dynamic equations describing the average behavior of the converter can be obtained as follows

$$L\frac{di}{dt} = -uv + v_i(t), \quad (1)$$

$$C\frac{dv}{dt} = ui - Gv, \quad (2)$$

where $i \in \mathbb{R}$ describes the current flows in the inductance $L$, and $v \in \mathbb{R}_{>0}$ is the voltage across both the capacitance $C$ and the load conductance $G$. The continuous signal $u \in [-1, 1]$ operates as a control input and is fed to the PWM circuit to generate the sequence of switching positions $\delta_1$ and $\delta_2$ their complements $\bar{\delta}_1$ and $\bar{\delta}_2$, respectively. The switch position function takes the values in finite set $\{0, 1\}$ and the exact model can be obtain by standing $\delta_1 - \delta_2$ as the control signal in (1), (2). Finally, $v_i(t) = E\sin(\omega t)$ represents the voltage of the AC-input.

In the system shown in Fig 1, the control objective is to regulate (in average) the output (capacitor) voltage in some constant desired value $V_d > E$ with a nearly unity power factor in input side. Due to its applications such as interfacing renewable energy sources to hybrid microgrids [1], flexible AC transmission systems [2], motor drive systems [3], LED drive systems [4], many studies have been done to develop advanced strategy to meet the mentioned objectives.

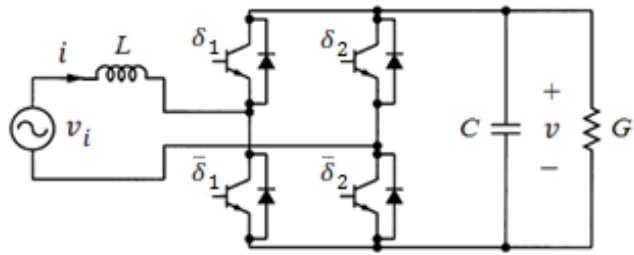

Figure 1.  AC-DC full-bridge boost converter circuit [5].

In the problem of PFC of an AC-DC boost converter, knowledge about the amplitude of AC-input and load conductance is vital to provide the feedforward control signal; see control signals in [6] and [7]. Besides an accurate measuring of these parameters and feedback from input current improves the output voltage regulation, total harmonic distortion (THD), and power factor quality [6].

It is worth noting that, the AC-DC boost converter belongs to second-order bilinear systems group and both of its states are not observable when the control signal is zero. These features pose an interesting state and parameter estimating problem which also bring down the number of sensors and costs in industrial applications. Since, an asymptotic and classical current observer is designed for the interleaved and non-interleaved AC-DC boost converters in [3, 4, 8]. Indeed, a sliding mode observer is proposed in [9] with exponential current estimation for the three phase topology of the system (1), (2). All the current observers proposed in [3, 4, 8] and [9] require to measure the amplitude of the AC-source, addition to the knowledge about the load conductance.

The main contribution of this paper is to design a nonlinear, globally convergent, and robust estimator via immersion and invariance (I&I) based filtered transformation. In the proposed method, the input current, load conductance, and input voltage amplitude are globally exponentially estimated from the output voltage via a fifth-dimensional estimator. An application of the estimator is presented in conjunction with a well-known dynamic controller.

## II. PROBLEM STATEMENT

Let us to assume that the steady-state form the input current of the system (1), (2) can be written as follows

$$i_s(t) = I_s \sin(\omega t - \Delta\rho), \quad (3)$$

M. Tavan is with the Department of Electrical Engineering, Mahmudabad Branch, Islamic Azad University, Mahmudabad, Iran (e-mail: m.tavan@srbiau.ac.ir).
K. Sabahi is with the Department of Electrical Engineering, Mamaghan Branch, Islamic Azad University, Mamaghan, Iran.

where $\Delta\rho$ is the phase difference between input voltage and current, and $I_s \in \mathbb{R}_{>0}$ yet to be specified. Replacing (3) into (1) yields

$$u_s v_s = E\sin(\omega t) - L\omega I_s \cos(\omega t - \Delta\rho), \quad (4)$$

where $u_s$ and $v_s$ are the control input and output voltage in steady-state, respectively. Now, replacing (3) and (4) into (2) yields the following dynamics in the steady-state

$$Cv_s \dot{v}_s + Gv_s^2 = EI_s \sin(\omega t)\sin(\omega t - \Delta\rho) - \frac{L\omega I_s^2}{2}\cos(2\omega t - 2\Delta\rho). \quad (5)$$

The steady-state solution of (4) can be calculated using Fourier series and is given by

$$v_s^2(t) = \frac{EI_s}{2G}\cos(\Delta\rho) - \frac{Gd_1 - C\omega d_2}{G^2 + C^2\omega^2}\cos(\omega t) - \frac{Gd_2 + C\omega d_1}{G^2 + C^2\omega^2}\sin(\omega t), \quad (6)$$

with

$$d_1(\Delta\rho) := \frac{I_s}{2}(E\cos(\Delta\rho) - L\omega I_s \sin(2\Delta\rho)), \quad (7)$$

$$d_2(\Delta\rho) := \frac{I_s}{2}(E\sin(\Delta\rho) + L\omega I_s \cos(2\Delta\rho)). \quad (8)$$

Finally, doing a basic trigonometric simplification (6) takes the form

$$v_s^2(t) = \frac{EI_s}{2G}\cos(\Delta\rho) + A\sin(2\omega t + \varepsilon), \quad (9)$$

with

$$A(\Delta\rho) := \sqrt{\frac{d_1^2 + d_2^2}{G^2 + C^2\omega^2}}, \quad (10)$$

$$\varepsilon(\Delta\rho) := \arctan 2\left(\frac{Gd_1 - C\omega d_2}{Gd_2 + C\omega d_1}\right). \quad (11)$$

The new representation, given by (9), demonstrates the effect of the phase shift on the DC value and the amplitude of the second-order harmonic of the output in steady-state.

*Remark 1:* The DC-component of $v_s$ in steady state can be concluded from (9) and is given by

$$V_s(\Delta\rho, I_s) = \sqrt{\frac{EI_s}{2G}\cos(\Delta\rho)}. \quad (12)$$

As a result, in order to provide the desired output voltage $V_d$, the input current amplitude must be forced to achieve

$$I_s(\Delta\rho) = \frac{2GV_d^2}{E\cos(\Delta\rho)}, \quad (13)$$

where for $\Delta\rho = 0$, the minimum value of $I_s$ can be achieved and is defined as

$$I_0 := \frac{2GV_d^2}{E}, \quad (14)$$

which yields to maximum efficiency.

*Remark 2:* The condition $v \in \mathbb{R}_{>0}$ implies $V_s(\Delta\rho, I_s) \in \mathbb{R}_{>0}$. In turn, with respect to (12) this condition implies that $\Delta\rho \in (-\pi/2, \pi/2)$. Note now that (4) can be simplified as

$$u_s v_s = B\sin(\omega t + \delta), \quad (15)$$

with

$$B(\Delta\rho) := \sqrt{E^2 + L^2\omega^2 I_s^2 + 2EL\omega I_s \sin(\Delta\rho)}, \quad (16)$$

$$\delta(\Delta\rho) := \arctan 2\left(\frac{L\omega I_s \cos(\Delta\rho)}{E + L\omega I_s \sin(\Delta\rho)}\right). \quad (17)$$

As a result, for any $\Delta\rho \in (-\pi/2, \pi/2)$ and positive value of $E$, $L$, $\omega$, and $I_s$, the term $u_s v_s$ remains a sinusoidal signal with non-zero amplitude and phase. In other word, the sign of the term $u_s v_s$ changes over time.

It is well-known that the system (1), (2) is nonminimum-phase associated to the output to be regulated. Hence, to circle this problem in most control algorithms the output voltage is regulated by means of indirectly stabilizing the input current to its desired value [6, 10]. On the other hand, from the results of Remark 1, the desired input current amplitude depends upon the values of the load conductance and input voltage amplitude. Hence, in this paper we are interested to estimate the parameter

$$\theta := [E, G]^\top, \quad (18)$$

from the output voltage $v$. Note that, the system equations (1), (2) can be linearly parameterised in term of $\theta$ by the regressors

$$\varphi_1(t) := \frac{1}{L}[\sin(\omega t), 0]^\top, \quad (19)$$

$$\varphi_0(v) := \frac{1}{C}[0, -v]^\top, \quad (20)$$

as

$$\frac{di}{dt} = -\frac{u}{L}v + \varphi_1^\top \theta, \quad (21)$$

$$\frac{dv}{dt} = \frac{u}{C}i + \varphi_0^\top \theta. \quad (22)$$

III. ESTIMATOR DESIGN

In this section an estimator is designed for the system (1), (2) which requires the knowledge about $\omega$, $C$, and $L$. It should be noted that the only measurable state is the output voltage $v$. This cancels the use for input current sensor and makes the practical implementation of the proposed procedure attractive.

A. Design procedure

In order to form a proper adaptive structure the following input-output filtered transformation is considered

$$\iota(t) := i(t) - \mu^\top(t)\theta, \quad (23)$$

where $\mu: \mathbb{R}_+ \to \mathbb{R}^2$ is an auxiliary dynamic vector, which its dynamics are to be defined. The equation above admits the global inverse

$$i = \iota + \mu^\top \theta$$
$$= [1 \quad \mu^\top]\eta, \tag{24}$$

where
$$\eta := \operatorname{col}(\iota, \theta) \tag{25}$$

is the new unavailable vector. With respect to (21), the dynamics of $\eta$ can be obtain as

$$\dot{\eta} = \mathcal{A}(t)\eta - \frac{u}{L}\begin{bmatrix} v \\ \sigma_2 \end{bmatrix}, \tag{26}$$

with
$$\mathcal{A}(t) := \begin{bmatrix} 0 & \varphi_1^\top - \dot{\mu}^\top \\ \sigma_2 & O_2 \end{bmatrix}, \tag{27}$$

where $O_2 \in \mathbb{R}^{2\times 2}$ and $\sigma_2 \in \mathbb{R}^2$ are the zero matrix and zero vector, respectively. Also, the dynamics of the output voltage (22) can be rewritten as

$$\dot{v} = \mathcal{C}^\top(v, u, \mu)\eta \tag{28}$$

with
$$\mathcal{C}^\top(v, u, \mu) := \frac{1}{C}[u \quad u\mu^\top + C\varphi_0^\top]$$
$$=: [\mathcal{C}_1(u) \quad \mathcal{C}_2^\top(v, u, \mu)]. \tag{29}$$

Now, under the inspiration of the adaptive I&I technique [5], let us to define the estimation error as

$$\bar{\eta} = \eta - \zeta - \beta(v, t), \tag{30}$$

where $\zeta \in \mathbb{R}^3$ is the estimator state, and the mapping $\beta: \mathbb{R}_{>0} \times \mathbb{R}_+ \to \mathbb{R}^3$ yet to be specified. To obtain the dynamics of $\bar{\eta}$ differentiate (30), yielding

$$\dot{\bar{\eta}} = \left(\mathcal{A} - \frac{\partial \beta}{\partial v}\mathcal{C}^\top\right)\eta - \frac{u}{L}\begin{bmatrix} v \\ \sigma_2 \end{bmatrix} - \dot{\zeta} - \frac{\partial \beta}{\partial t}, \tag{31}$$

which use has been made of (26) and (27). Assume that the last term in the right hand side of the above equation is known, then selecting the update law as

$$\dot{\zeta} = \left(\mathcal{A} - \frac{\partial \beta}{\partial v}\mathcal{C}^\top\right)(\zeta + \beta) - \frac{u}{L}\begin{bmatrix} v \\ \sigma_2 \end{bmatrix} - \frac{\partial \beta}{\partial t}, \tag{32}$$

cancels the known terms in the error dynamics and yields to

$$\dot{\bar{\eta}} = \left(\mathcal{A} - \frac{\partial \beta}{\partial v}\mathcal{C}^\top\right)\bar{\eta}. \tag{33}$$

It is clear that the convergence of the proposed estimator hinges upon to select of a proper dynamic for $\beta$ and $\mu$. Our suggestion for $\beta$ is

$$\beta(v, t) = k\begin{bmatrix} 1 & \sigma_2^\top \\ \sigma_2 & \mathcal{T}\mathcal{D} \end{bmatrix}\int_0^v \mathcal{C}(y, u, \mu)dy$$
$$= k\begin{bmatrix} \int_0^v \mathcal{C}_1(u)dy \\ \mathcal{T}\mathcal{D}\int_0^v \mathcal{C}_2(y, u, \mu)dy \end{bmatrix} =: \begin{bmatrix} \beta_1(v, u) \\ \beta_2(v, u, \mu) \end{bmatrix}, \tag{34}$$

where $\mathcal{T} = \mathcal{T}^\top$ and $\mathcal{D} = \mathcal{D}^\top$ are constant and positive definite matrixes, and $k \in \mathbb{R}_{>0}$ is a constant value. The proposed mapping makes positive-semidefinite the sign of the matrix $\frac{\partial \beta}{\partial v}\mathcal{C}^\top$. Also, picking $\mu$-dynamics as

$$\dot{\mu} = -\left(\mathcal{C}_2\frac{\partial \beta_1}{\partial v} + \mathcal{T}^{-1}\frac{\partial \beta_2}{\partial v}\mathcal{C}_1\right) + \varphi_1$$
$$= -k(\mathcal{I}_2 + \mathcal{D})\mathcal{C}_2\mathcal{C}_1 + \varphi_1, \tag{35}$$

where $\mathcal{I}_2$ is the $2 \times 2$ identity matrix, enforces the following dynamics along the error trajectories

$$\dot{\bar{\eta}} = -k\begin{bmatrix} \mathcal{C}_1^2 & -\mathcal{C}_1\mathcal{C}_2^\top\mathcal{D} \\ \mathcal{T}\mathcal{D}\mathcal{C}_2\mathcal{C}_1 & \mathcal{T}\mathcal{D}\mathcal{C}_2\mathcal{C}_2^\top \end{bmatrix}\bar{\eta}. \tag{36}$$

### B. Stability analysis

To indicate the dynamics characteristics of the system above consider the Lyapunov function

$$V(\bar{\eta}) = \frac{1}{2}\bar{\eta}^\top\begin{bmatrix} 1 & \sigma_2^\top \\ \sigma_2 & \mathcal{T}^{-1} \end{bmatrix}\bar{\eta}, \tag{37}$$

whose time-derivative along the trajectories (36) is given by

$$\dot{V}(\bar{\eta}) = -k\bar{\eta}^\top\begin{bmatrix} \mathcal{C}_1^2 & \sigma_2^\top \\ \sigma_2 & \mathcal{D}\mathcal{C}_2\mathcal{C}_2^\top \end{bmatrix}\bar{\eta}$$
$$=: \bar{\eta}^\top Q(t)\bar{\eta}. \tag{38}$$

Note that $Q(t)$ is a positive semi-definite matrix. Therefore, the origin of the system is a uniformly stable equilibrium point and the estimation error $\bar{\eta}$ is bounded for all times. Also, exploiting the proof of Proposition 1 in [11] implies that the zero equilibrium of the system (36) is globally exponentially stable if $\mathcal{C}_1$ and $\mathcal{C}_2$ satisfy the persistency of excitation (PE) condition.

It is well-known that $\mathcal{C}_1 \in$ PE or equivalently $u \in$ PE is satisfied when the bounded signal $u$ is nonzero in steady state. This condition is ensured in the operation mode in which the system is forced to track a positive constant as the desired output. On the other hands, the vector regressor $\mathcal{C}_2 \in \mathbb{R}^2$ is PE if there exist a constant $T_0 \in \mathbb{R}_{>0}$ such that

$$\infty > \int_t^{t+T_0}\left[\mathcal{C}_2^\top(v(\tau), u(\tau), \mu(\tau))\epsilon_0\right]^2 d\tau > 0, \tag{39}$$

for any $t \in \mathbb{R}_+$, and all vector $\epsilon_0 \in \mathbb{R}^2$ with $|\epsilon_0| = 1$ [12]. As a result, $\mathcal{C}_2$ is PE if its entries are also PE and linearly independent. To this end, we need first to establish boundedness of $\mathcal{C}_2$ and $\dot{\mathcal{C}}_2$.

In real application, the AC-source has a positive internal resistance. Therefore, from Lemma 8.1 in [6] it can be concluded that the system states $v(t)$ and $i(t)$ and their dynamics remain bounded for any bounded input voltage. Now, without loss of generality assume $\mathcal{D} = \operatorname{diag}(d_1, d_2)$ then substituting (19), (20), and (29) in (35) yields the asymptotically stable filters

$$\dot{\mu}_1 = -k(1 + d_1)\left(\frac{u}{C}\right)^2\mu_1 + \frac{1}{L}\sin(\omega t), \tag{40}$$

$$\dot{\mu}_2 = -k(1 + d_2)\frac{u}{C}\left(\frac{u}{C}\mu_2 - \frac{1}{C}v\right), \tag{41}$$

where $\operatorname{col}(\mu_1, \mu_2) = \mu$. Due to boundedness of the filters inputs, $\dot{\mu}$ and $\mu$ are bounded. Then boundedness of the regressor $\mathcal{C}_2$ and $\dot{\mathcal{C}}_2$ can be concluded.

From (29), the first entry of $\mathcal{C}_2$ is given by $\mathcal{C}_{21} = C^{-1}u\mu_1$. Since, the PE property is preserved under filtering by an asymptotic stable dynamics, then with respect to (40),

$u\mu_1 \in$ PE if $u\sin(\omega t) \in$ PE which is guaranteed invoking Proposition 2 in [13]. The second entry of $\mathcal{C}_2$ is given by $\mathcal{C}_{22} = C^{-1}(u\mu_2 - v)$. The PE property fails for $\mathcal{C}_{22}$ if $u\mu_2 = v$ in steady state. The impossibility of this condition is established by contradiction with respect to Remark 2. Invoking the equality in (41) leads to $\dot{\mu}_2 = 0$ and then a constant value for $\mu_2$. Also, from the equality we get $uv = \mu_2 u^2$. This implies a unique sign for the term $uv$ and in turn for $u_s v_s$ which is in contradiction to (15) and its results in Remark 2. Then $\mathcal{C}_{22}$ is a non-zero signal in steady state and then $\mathcal{C}_{22} \in$ PE. Finally, $\mathcal{C}_2$ is PE vector regressor if $\mathcal{C}_{21}$ and $\mathcal{C}_{22}$ be linearly independent in steady state. Once again, we equivalently show the property by contradiction, i.e. does not exist a constant value $p \in \mathbb{R}$ such that $u\sin(\omega t) = p(u^2+1)v$. Multiplying both side by $v$ yields $uv\sin(\omega t) = p(u^2+1)v^2$. This implies a unique sign for $uv\sin(\omega t)$ and in turn for $u_s v_s \sin(\omega t) = B\sin(\omega t + \delta)\sin(\omega t)$ which is not true for $\delta \neq 0$ such as given by (17). Therefore the PE property is guaranteed for $\mathcal{C}_2$ and the estimation error $\bar{\eta}$ converges to zero.

### C. State and parameters estimation

Invoking the attractivity of the equilibrium point $\bar{\eta} = 0$ in (30) implies an asymptotically converging estimate of $\eta$ given by

$$\hat{\eta} = \zeta + \beta(v,t). \tag{42}$$

From (25), (34), and the above equation the parameter vector estimation can be derived as

$$\hat{\theta} = \zeta_2 + k\mathcal{T}\mathcal{D}\left(u\mu - \frac{1}{2}\begin{bmatrix}0\\v\end{bmatrix}\right)\frac{v}{C}, \tag{43}$$

where $\zeta_2 \in \mathbb{R}^2$ is the corresponding estimator state. Also, with respect to (24) an estimate of the input current can be obtained by

$$\hat{\imath} = \zeta_1 + ku\frac{v}{C} + \mu^\top \hat{\theta}, \tag{44}$$

where $\text{col}(\zeta_1, \zeta_2) = \zeta$.

## IV. ADAPTIVE OUTPUT FEEDBACK CONTROL

The update law of the proposed estimator need to the dynamics of control signal. Hence, the dynamic controller introduced in Proposition 8.9 of [6], is suggested in conjunction with the estimator. For the dynamic model given by (1) and (2), the control law can be rewritten as [5]

$$\dot{u} = \frac{1}{v}\left(-\frac{u^2}{C}i + w\right), \tag{45}$$

where $w(t)$ is derived from the filter

$$w(s) = c\frac{s^2 + as + b}{s^2 + \omega}e(s), \tag{46}$$

with some positive constants $a$, $b$, $c$, and

$$e := v_i - L\frac{di_d}{dt} - k(i_d - i) - uv, \tag{47}$$

with positive constant value $k$, where $i_d = I_0 \sin(\omega t)$ and $I_0$ is given by (14). For sufficiently large $c$, the closed loop system is asymptotically stable, and then $e$ converges to zero and $i$ converges to $i_d$ [6]. On the other hand, from the analysis in Remark 1, for $\Delta \rho = 0$ and $I_s = I_0$ in (12), the output voltage $v$ converges in average to $V_d$. Note that, $v_i$ and $i_d$ in (47) are sinusoid signals, hence we can conclude that $u$ is sinusoid in the steady state. As a result the assumption $u \in$ PE is satisfied.

The feedback term of the above control law needs an accurate measuring of the input current, whereas *a priori* knowledge of the load conductance and input voltage amplitude is needed for the feedforward term. Hence, a *certainty equivalent* version of the controller, i.e., replacing $\theta = \text{col}(E,G)$ and $i$ by their estimation given by (43) and (44), respectively, is suggested with some modification in control law.

Notice that, with respect to (14), singularity problem can be occurred when $\hat{E}$ is used to estimate $I_0$ in $i_d$. In order to circumvent the problem, we assume that $E^{-1}$ is lumped in the arbitrary gain $c$ in (46) and the signal

$$\bar{e} := Ee, \tag{48}$$

is fed to the filter (46) instead of $e$. An estimate of $\bar{e}$ can be obtained by

$$\hat{\bar{e}}(t) = \hat{q}\cos(\omega t - \hat{p}) - \hat{E}(uv - k\hat{\imath}), \tag{49}$$

with

$$\hat{q} := \sqrt{\left(\hat{E}^2 - 2k\hat{G}V_d^2\right)^2 + \left(2\hat{G}V_d^2 L\omega\right)^2}, \tag{50}$$

$$\hat{p} := \arctan\left(\frac{\hat{E}^2 - 2k\hat{G}V_d^2}{-2\hat{G}V_d^2 L\omega}\right). \tag{51}$$

Finally, the control system in the closed loop with the estimator will be complete by

$$\dot{u} = \frac{1}{v}\left(-\frac{u^2}{C}\hat{\imath} + \widehat{w}\right), \tag{52}$$

with

$$\widehat{w}(s) = d\frac{s^2 + as + b}{s^2 + \omega}\hat{\bar{e}}(s), \tag{53}$$

where $d$ is related to $c$ in (46) with $d = c/E$.

*Remark 4:* Individually estimation of the AC-side parasitic resistance proposed in [6] is extra when an estimation of input voltage amplitude is making. Note that, its effect can be added in (47) as a small voltage drop in phase with the AC-source (see equation (23) in [6]). Therefore, the incorporation of this uncertainty into the estimated amplitude of input voltage is done in the modified control law above.

## V. CONCLUSION

A new nonlinear, globally exponentially convergent, robust estimator is designed via I&I based filtered transformation. The input current, load conductance and the input voltage amplitude are estimated from the output voltage via a fifth-dimensional estimator. Also, an application of the estimator is presented in conjunction with a well-known dynamic controller.

## REFERENCES

[1] X. Lu, J. M. Guerrero, K. Sun, J. C. Vasquez, R. Teodorescu, and L. Huang, "Hierarchical control of parallel AC-DC converter interfaces for hybrid microgrids," *IEEE Transactions on Smart Grid,* vol. 5, pp. 683-692, 2014.

[2] A. M. Bouzid, J. M. Guerrero, A. Cheriti, M. Bouhamida, P. Sicard, and M. Benghanem, "A survey on control of electric power distributed generation systems for microgrid applications," *Renewable and Sustainable Energy Reviews,* vol. 44, pp. 751-766, 2015.

[3] G. Cimini, M. L. Corradini, G. Ippoliti, G. Orlando, and M. Pirro, "Passivity-based PFC for interleaved boost converter of PMSM drives," *IFAC Proceedings Volumes,* vol. 46, pp. 128-133, 2013.

[4] G. Calisse, G. Cimini, L. Colombo, A. Freddi, G. Ippoliti, A. Monteriù*, et al.*, "Development of a smart LED lighting system: Rapid prototyping scenario," in *Systems, Signals & Devices (SSD), 2014 11th International Multi-Conference on*, 2014, pp. 1-6.

[5] A. Astolfi, D. Karagiannis, and R. Ortega, *Nonlinear and adaptive control with applications*: Springer Science & Business Media, 2007.

[6] D. Karagiannis, E. Mendes, A. Astolfi, and R. Ortega, "An experimental comparison of several PWM controllers for a single-phase AC-DC converter," *IEEE Transactions on control systems technology,* vol. 11, pp. 940-947, 2003.

[7] R. Cisneros, M. Pirro, G. Bergna, R. Ortega, G. Ippoliti, and M. Molinas, "Global tracking passivity-based pi control of bilinear systems: Application to the interleaved boost and modular multilevel converters," *Control Engineering Practice,* vol. 43, pp. 109-119, 2015.

[8] G. Cimini, G. Ippoliti, G. Orlando, and M. Pirro, "Sensorless power factor control for mixed conduction mode boost converter using passivity-based control," *IET Power Electronics,* vol. 7, pp. 2988-2995, 2014.

[9] J. Liu, S. Laghrouche, and M. Wack, "Observer-based higher order sliding mode control of power factor in three-phase AC/DC converter for hybrid electric vehicle applications," *International Journal of Control,* vol. 87, pp. 1117-1130, 2014.

[10] G. Escobar, D. Chevreau, R. Ortega, and E. Mendes, "An adaptive passivity-based controller for a unity power factor rectifier," *IEEE Transactions on Control Systems Technology,* vol. 9, pp. 637-644, 2001.

[11] M. Tavan and K. Sabahi, "Input Voltage and Current Sensorless Control of a Single Phase AC-DC Boost Converter," *arXiv preprint arXiv:1804.00342,* 2018.

[12] S. Sastry and M. Bodson, *Adaptive control: stability, convergence and robustness*: Courier Corporation, 2011.

[13] A. A. Bobtsov, A. A. Pyrkin, R. Ortega, S. N. Vukosavic, A. M. Stankovic, and E. V. Panteley, "A robust globally convergent position observer for the permanent magnet synchronous motor," *Automatica,* vol. 61, pp. 47-54, 2015.